\documentclass{osa-article}
\usepackage{graphicx}
\journal{oe}

\newcommand{\eref}[1]{Eq.~(\ref{#1})}
\newcommand{\fref}[1]{Fig.~\ref{#1}}

\newcommand{\sref}[1]{Sec.~\ref{#1}}

\newcommand{\df}{\ensuremath{\mathrm{d}}}

\newcommand{\calM}{\ensuremath{\mathcal{M}}}

\newcommand{\ie}{\emph{i.e.}}

\newcommand{\etc}{\emph{etc.}}
\newcommand{\cd}{\ensuremath{\mathrm{C}^\circ}}

\begin{document}

\title{Why are thermal images blurry}

\author{Fanglin Bao,\authormark{1,3} Shubhankar Jape,\authormark{1} Andrew Schramka,\authormark{1} Junjie Wang, \authormark{2} Tim E. McGraw, \authormark{2} and Zubin Jacob \authormark{1,4}}

\address{\authormark{1}Birck Nanotechnology Center, School of Electrical and Computer Engineering, Purdue University, West Lafayette, IN 47907, USA\\
\authormark{2}Department of Computer Graphics Technology, Purdue University, West Lafayette, IN 47907, USA\\
\authormark{3}baof@purdue.edu\\
\authormark{4}zjacob@purdue.edu}



\begin{abstract*}
The resolution of optical imaging is limited by diffraction as well as detector noise. However, thermal imaging exhibits an additional unique phenomenon of ghosting which results in blurry and low-texture images. Here, we provide a detailed view of thermal physics-driven texture and explain why it vanishes in thermal images capturing heat radiation. We show that spectral resolution in thermal imagery can help recover this texture, and we provide algorithms to recover texture close to the ground truth. Using a simulator for complex 3D scenes, we discuss the interplay of geometric textures and non-uniform temperatures which is common in real-world thermal imaging. We demonstrate the failure of traditional thermal imaging to recover ground truth in multiple scenarios while our thermal perception approach successfully recovers geometric textures. Finally, we put forth an experimentally feasible infrared Bayer-filter approach to achieve thermal perception in pitch darkness as vivid as optical imagery in broad daylight.
\end{abstract*}

\section{Introduction}
Thermal imaging has myriad applications from surveillance and defense to advanced driver-assistance systems and autonomous navigation \cite{Gade2014}.
Originally developed as a night-vision tool for defense applications, thermal imaging has now become a versatile consumer technology due to its ability to see through bad weather \cite{9133581}, darkness \cite{s16060820}, or act as a thermography technique for biomedical applications \cite{Tang2020}. According to Planck's law, every object at finite temperature $T$, including human bodies, ground, and buildings, would emit infrared thermal radiation, depending on the object's emissivity $e$. Thermal radiation propagates and scatters off multiple objects, and is omnipresent in both day and night. This sensing modality works in the mid-wave and long-wave infrared spectrum and is complementary to visible-light optical imaging. Nevertheless, thermal images are known to be of low contrast and lack details and are often blurry compared to optical images \cite{Gurton2014,Treible_2017_CVPR}. This phenomenon of blurry thermal images is partly attributed to the sensor noise, lower pixel array sizes as well as the wavelength differences between infrared and optical radiation. However, the goal of this paper is to show that a fundamental mechanism is at play beyond the hardware differences between visible-light and thermal-infrared cameras.

Machine learning especially for visible-light optical images has led to revolutionary advances in machine perception \cite{geiger2012we}. On the other hand, the advances in the infrared thermal domain have been hindered due to the above-mentioned blurry nature of thermal images. For example, thermal imagery is often used only as a subsidiary sensing approach to enhance RGB images under poor ambient illumination \cite{s16060820}. Recently, it was shown that embedding the physics of thermal radiation in machine learning algorithms can overcome the blurry nature of thermal images \cite{Bao2023}. It was shown that machine perception using infrared thermal radiation can lead to imagery in pitch darkness as vivid as broad daylight \cite{Bao2023}. Thus a thermal perception technique like HADAR: Heat-Assisted Detection and Ranging \cite{Bao2023} can compete with visible-light optical imaging as a stand-alone sensing modality in the future. However, open challenges remain in analyzing complex scenes with non-uniform temperature distributions. Furthermore, the robustness of the inverse estimation problem in multiple scenarios with resource constraints remains untested.

In this paper, our first goal is to explain the underlying mechanism that causes thermal images to lose texture and become blurry. This mechanism which is called `ghosting' is unique to infrared thermal imaging. It does not occur in visible-light optical imaging. Furthermore, the limitations in texture/resolution are distinct from the well-known diffraction limit as well as the noise limit arising from detector non-idealities. We develop a thermal simulator specifically for three-dimensional real-world scenes with non-uniform temperature distribution. We show that thermal physics-driven perception algorithms can correctly recover the geometric textures of realistic scenes with non-uniform temperatures. Unlike previous work \cite{Bao2023}, we do not restrict ourselves to scenes with uniform temperatures. Through extensive numerical simulations, we shed light on the thermal physics-driven definition of texture. This can have important applications for designing algorithms where physics can be embedded into the signal processing and machine learning approaches.

The necessary hardware modification to overcome ghosting is based on the spectral resolution of thermal images. Spectral resolution increases the complexity of the thermal cameras but leads to higher texture recovery in the collected images. Here, we demonstrate the role of the number of spectral bands in recovering the ground truth of complex scenes using our thermal simulator. We show that the error in estimating the scene decreases as the number of spectral bands increases.  Long-wave infrared hyperspectral imaging underlying the TeX vision theory \cite{Bao2023} is experimentally challenging. Here, TeX vision stands for the representation of heat radiation where the pixels in images are represented by three physical attributes of temperature ($T$), emissivity ($e$), and physics-driven texture ($X$). We demonstrate that we can achieve a TeX vision representation close to the ground truth with the Bayer-filter approach (using only 4 Bayer filters). Similar to optical frequency Bayer filters, we believe our design can have widespread industrial applications. Our results can lead to the design of optimal spectral modules for next-generation thermal imagers.

We show that thermal-physics driven perception has significant advantages in performance over existing state-of-the-art physics-agnostic approaches. Our paper uses the Contrast-Limited Adaptive Histogram Equalization (CLAHE) \cite{Khare2021} as the state-of-the-art baseline of image processing. Among a variety of digital image processing algorithms for contrast enhancement of thermal images \cite{Soundrapandiyan2022}, histogram equalization and its variants \cite{Khare2021,Dhal2021,LI2018164} are widely adopted techniques in state-of-the-art thermal datasets such as the FLIR thermal dataset. Machine-learning-based approaches \cite{Bouhlel2023,Pang2023,KUANG2019119,Lee2017} also present a new research frontier in improving the visual contrast by learning the multi-scale thermal features. Those image processing algorithms aim to recover the scattering signal by heuristically removing the strong contribution of direct emission. Our work shows that image processing fails to recover geometric textures when the direct emission is spatially non-uniform. We argue that this is the fundamental reason causing the persistence of the ghosting effect in thermal imaging in spite of significant advances in camera hardware (low-noise cooled sensors, larger pixel array sizes \etc).

This paper is organized as follows. \sref{sec:tex} shows general examples to explain the loss of geometric textures and the emergence of the ghosting effect. \sref{sec:gttex} briefly introduces the TeX vision for completeness. \sref{sec:sgd} investigates TeX-SGD (semi-global decomposition) with non-uniform temperature and demonstrates texture recovery close to the ground truth. \sref{sec:tmp} shows the main results of this paper on the interplay of geometric textures and non-uniform temperature. This section also demonstrates that TeX vision beats existing techniques in overcoming the ghosting effect. \sref{sec:res} and \sref{sec:env} show the main results on the influence of spectral resolution and finite cutoff of environmental objects in TeX vision.

\section{Loss of geometric texture in thermal imaging}\label{sec:tex}
The mechanism of the ghosting effect can be revealed by analogy to a shining bulb \cite{Bao2023}. On a glowing bulb, it is impossible to identify the surface features, spot the geometric texture, or read the written text. However, as soon as it is turned off, the radiation from other sources scatters off the surface of the bulb rendering the surface features visible to the human eye or a camera. The key aspect is that the intrinsic radiation from the object mixes with the extrinsic scattering from nearby objects leading to loss of information/texture. Following the theory of thermal radiation and scattering, the total thermal radiation, or heat signal, $S$, entering a thermal camera is given by \cite{Bao2023},
\begin{equation}\label{eq:heatsignal}
    S_{\alpha\nu} = e_{\alpha\nu}B_{\nu}(T_{\alpha}) +[1- e_{\alpha\nu}] X_{\alpha\nu},
\end{equation}
with 
\begin{equation}\label{eq:x}
    X_{\alpha\nu} =  \sum_{\beta\neq\alpha} V_{\alpha\beta}S_{\beta\nu},
\end{equation}
where $B$ is the blackbody radiation given by Planck's law, $V$ is the thermal lighting factor related to the object's geometric surface normals, $\alpha$ is the object index, and $\nu$ is the wavenumber. Here, we assume a Lambertian emitter but our equations can be generalized to non-Lambertian emitters as well. The crucial thermal variable carrying texture information is $X$ which we define as the physics-driven geometric texture. The lack of geometric texture, \ie, lack of 3D surface normal information is the loss of information in the thermal variable $X$. This is because the direct emission term (first term) often dominates in the infrared spectral range. Geometric texture in imagery is crucial for machine perception tasks such as ranging and semantic segmentation. Thus a major goal of our work is to recover the texture $X$ using advanced thermal perception algorithms and custom sensors.

For a shining bulb or natural objects with near-unity emissivity ($e\approx 1$), blackbody radiation dominates the total heat signal, that is, $S_{\alpha\nu} \approx B_{\nu}(T_{\alpha})$, according to \eref{eq:heatsignal}. This explains why thermal imaging is widely utilized for measuring the temperature $T$ using primarily the intrinsic radiation from the object of interest. However, we emphasize that no information on the object's geometric textures is revealed in this direct emission. As explained before, the geometric textures on a shining bulb can only be seen by turning the bulb off. However, the fundamental blackbody radiation from natural objects can never be turned off leading to the ghosting effect. For a panchromatic thermal image, $S=\int S_\nu\,\df\nu$, separating the intrinsic and extrinsic signals is extremely challenging or impossible. The texture $X$ is irreversibly lost in the broadband image since there exist infinite solutions of $T$, $e_\nu$, and $X_\nu$, which can lead to the same observed signal. We address this as TeX degeneracy of panchromatic thermal imagery \cite{Bao2023}. We note that post-processing algorithms cannot isolate the direct emission term, $e\cdot B$, from the total signal  $S$ which causes the ghosting effect. This TeX degeneracy is the (novel) mechanism that causes the ghosting effect and is in stark contrast to well-known factors of diffraction and sensor noise.

The insight of \eref{eq:heatsignal} is that geometry information does lurk inside the total heat signal as a weak contribution of magnitude order $1-e$, in a low-contrast background of magnitude order $e$. Note, here we address the direct emission as the background, \ie, the direct emission signal which is widely used in thermography is in fact the background as it does not contain texture. Based on inverse computational algorithms and spectral resolution, Ref.~\cite{Bao2023} shows that artificial intelligence (AI) using TeX vision can recover geometric textures from the heat signal and see through the pitch darkness like broad daylight, making the long-standing dichotomy between day and night for human beings obsolete.

\begin{figure}[htb]
\centering
\includegraphics{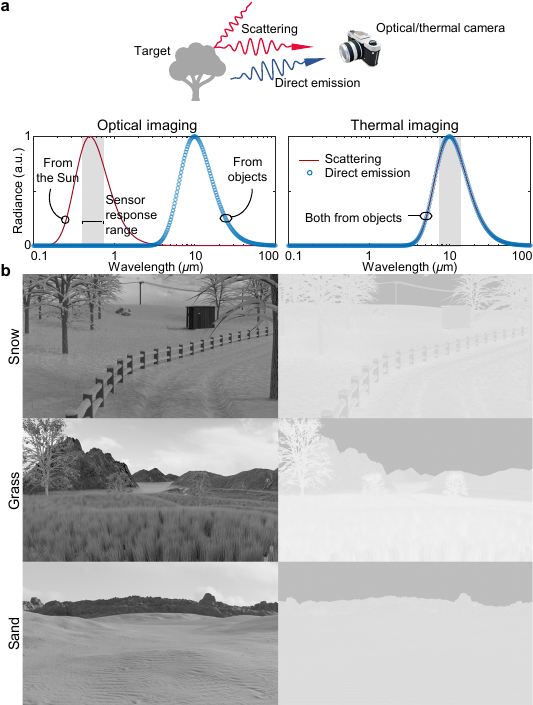}
\caption{The ghosting effect occurs when the geometric textures in the scattering signal are immersed in the strong direct emission of low contrast. (a) Typical radiance signal (normalized) for optical and thermal imaging. In optical imaging, the scattering signal (red curve) from solar illumination is well separated from the direct emission (blue circles) of scene objects, due to the high temperature contrast of the sun and our living environment. However, in thermal imaging, direct emission from targets largely overlaps with the scattering signal from environmental objects in the spectral domain, leading to the ghosting effect. Shaded areas are typical spectral response ranges for optical/thermal cameras. (b) Monte Carlo path tracing simulation of typical thermal (optical) imaging examples without (with) textures illustrating the ghosting effect. Optical and thermal imaging are set to have the same spatial resolution and sensor noise. Snow, grass, and sand are modeled as uniform materials at uniform temperatures but with different geometric surface-normal textures.}
\label{fig:tex}
\end{figure}

We note that \eref{eq:heatsignal} is a unified theory of optical imaging in daylight under solar illumination and thermal imaging at night. The differences between optical RGB imaging and long-wave infrared (LWIR) thermal imaging are roughly twofold, within the scope of this paper. Firstly, the spectral response range of the sensor for optical or thermal imaging is in the visible light range and LWIR, respectively. Secondly, the scattering signal of optical imaging is mainly the solar radiance, while the scattering signal of thermal imaging is from room-temperature objects. There is a large temperature contrast between the sun ($T\approx 5,500\,\cd$) and the scene objects on earth ($T\approx 20\,\cd$). According to Planck's law, the blackbody radiation is given by
\begin{equation}
    B_\lambda(T) = \frac{2hc^2}{\lambda^5}\frac{1}{e^{hc/(\lambda k_\mathrm{B}T)}-1},
\end{equation}
where $\lambda=1/\nu$ is the wavelength, $h$ is Planck's constant, $c$ is the speed of light, and $k_\mathrm{B}$ is Boltzmann's constant. Solar radiance peaks in the visible-light range, while room-temperature objects' radiance peaks in the LWIR, as shown in \fref{fig:tex}a.
For optical imaging, the scattering of solar illumination is well separated from the direct thermal radiation of scene objects, due to the high temperature contrast. This gives optical imaging vivid geometric textures encoded in the texture term $X$, as can be seen in the left column of \fref{fig:tex}b. Here, optical imaging is set as grayscale for a fair comparison. For thermal imaging, scattering and direct emission are both from scene objects around room temperatures. The fact that a traditional thermal camera collects both weak scattering and strong direct emission leads to textureless thermal images and thus the ghosting effect. See the right column of \fref{fig:tex}b.

We emphasize that visible light and LWIR radiation have different wavelengths, distinct sensor pixel sizes, and separate noise performance. The resulting different spatial and signal resolutions for the two modalities partially account for the different textures that can be captured by optical and thermal cameras. However, the insight from \fref{fig:tex} is that thermal imaging suffers from the ghosting effect and is textureless even if it has the same spatial and signal resolutions as optical imaging. The second insight from \fref{fig:tex} is that the vivid textures of snow, grass, and sand in optical imaging come merely from the geometric surface normals, where their material and temperature have been set to be uniform. These insights imply the possibility of recovering vivid textures through the geometric surface normals by properly removing the direct emission from the total heat signal.

\section{TeX vision overcomes the dichotomy between day and night}\label{sec:gttex}
\begin{figure}[htb]
\centering
\includegraphics{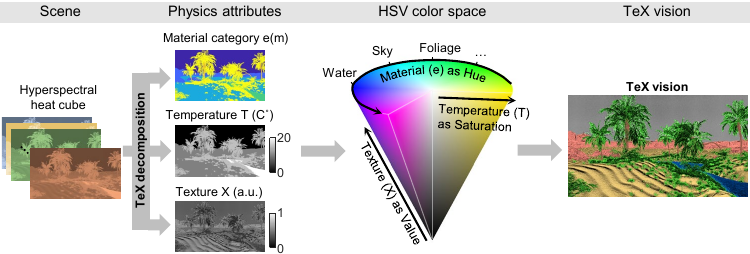}
\caption{Schematics of the TeX vision. TeX vision represents physics attributes of temperature $T$, material $e$, and texture $X$ in the HSV color space. Explicitly, color hue encodes the semantic material category, saturation encodes the temperature, and value/brightness encodes the texture. TeX vision enables AI to see through pitch darkness like broad daylight. The algorithm for TeX decomposition will be discussed in \sref{sec:sgd}.}
\label{fig:gttex}
\end{figure}

We emphasize that HADAR (Heat-Assisted Detection and Ranging) separates the scattering signal from the direct emission in thermal imaging using spectral resolution. To represent a hyperspectral imaging datacube, or the heat cube, traditional methods commonly use the principal component analysis (PCA) \cite{Du2007}. The PCA approach usually adopts the first 3 principal components to show in the RGB color space, while the rest components are discarded leading to information loss. In contrast, we note that the total information in the heat signal \eref{eq:heatsignal} can be captured in three physics quantities, namely, temperature $T$, emissivity $e$, and texture $X$. TeX vision decomposes these physics quantities from heat signal and shows them in the HSV color space, as illustrated in \fref{fig:gttex}. In the RGB color space, each channel represents a light intensity. In stark contrast, each channel in the HSV color space has a physical meaning. The Hue channel has a strong correlation with the daily semantics of objects. For example, blue usually indicates water or sky, green usually indicates grass or leaves, yellow usually indicates sand, \etc The Value/Brightness channel represents image textures of light and shadow. To reconstruct/mimic daylight RGB images with heat signals, in TeX vision, we show material category $e$ in the Hue channel and texture $X$ in the Value channel. For example, we assign to `water' a blue hue, `grass' a green hue, `sand' a yellow hue, and so on.

For TeX decomposition, we use a semantic library, $ \calM = \{e_{\nu}(m)|m = 1,2,...,M\}$, that approximates all possible spectral emissivities in the scene. For each object, its emissivity can be approximately described by one of the curves in the library, \ie, $e_{\alpha\nu} = e_{\nu}(m_\alpha)$. This semantic library can either be calibrated on-site or estimated from the heat cube itself \cite{Bao2023}. The adoption of a semantic library leads to the existence of a unique solution of the inverse TeX decomposition problem defined by \eref{eq:heatsignal}. The parameters to be estimated in the inverse problem are $\{T_\alpha, m_\alpha, V_{\alpha\beta}\}$. The detailed decomposition algorithm will be discussed in \sref{sec:sgd}. Since TeX vision records all the physics quantities, we argue that TeX vision is a full representation of the hyperspectral heat signal without losing information of any spectral bands. Furthermore, we emphasize that the colors in the TeX vision indicate material categories, $m_\alpha$, unlike pseudo coloring in traditional image processing. The accuracy of TeX decomposition depends on the accuracy of how the semantic library depicts the scene. An ideal semantic library becomes a ground-truth material library with exact spectral emissivity profiles. In this paper, we use the exact material library to demonstrate the TeX vision theory.

When the TeX decomposition in \fref{fig:gttex} is ideal, the resulting TeX vision strikingly shows an image of a night scene as if it is seen in daylight, with both recovered textures and semantic information. Artificial intelligence with TeX-based machine vision is thus able to overcome the long-standing dichotomy between day and night for human beings.

Note that the texture term $X$ in optical imaging usually involves only the sky and the sun as light sources. In thermal imaging, every object in the environment like streets and buildings has its scattering contributions in texture $X$. To mimic daylight optical imaging, scattering contributions of environmental objects other than the sky need to be removed. This process is discussed in \sref{sec:sgd}.

\section{TeX-SGD: Texture recovery close to the ground truth}\label{sec:sgd}
For each color pixel, TeX vision visualizes temperature $T$ as the saturation, material category $e(m)$ as the color hue, and texture $X$ as the brightness. By manually splitting between the scattering contribution and the direct emission contribution along with manually controlling the temperature and material in the Monte Carlo path tracing simulation, we can determine the ground truth TeX vision of a scene. The ground truth TeX vision in \fref{fig:gttex} shows the recovered textures of night scenes which are as vivid as when viewed in daylight. Now, we show how to recover the texture and generate the TeX vision close to the ground truth.

To tackle the iterative system of equations (\ref{eq:heatsignal}) and (\ref{eq:x}) with a possibly infinite number of environmental objects, we approximate the panoramic environment as $k$ equivalent environmental objects whose spectral emissivity are also among those $M$ curves in the semantic library $\calM$. The following reconstructed heat signal $\Tilde{S}^k_{\alpha\nu}$ with only $k$ environmental objects presents a good approximation of the original heat signal $S_{\alpha\nu}$,
\begin{equation}\label{eq:approxheat}
    \Tilde{S}^k_{\alpha\nu} = e_\nu(m_\alpha)B_{\nu}(T_{\alpha}) +[1- e_\nu(m_\alpha)] \Tilde{X}^k_{\alpha\nu},
\end{equation}
with 
\begin{equation}\label{eq:approxx}
    \Tilde{X}^k_{\alpha\nu} =  V_{\alpha 1}\Tilde{S}_{1\nu}+V_{\alpha 2}\Tilde{S}_{2\nu}+\cdots+V_{\alpha k}\Tilde{S}_{k\nu},
\end{equation}
where environmental radiance $\Tilde{S}_{1,\cdots,k;\nu}$ can be approximated from the captured images, see \sref{sec:env} for more details. The above approximation can be understood from the viewpoint of ray/path tracing. Path tracing of a real-world scene is asymptotically accurate and realistic, when the ray depth and meshing density of the environment increase.

TeX-SGD (semi-global decomposition) aims to extract the scene attributes, \ie, $T_\alpha$, $m_\alpha$, and $V_{\alpha;1,\cdots,k}$, by minimizing the $l_2$-norm residue, $\delta^k_\alpha \equiv ||\Tilde{S}^k_{\alpha\nu} - S_{\alpha\nu}||$,
\begin{equation}\label{eq:sgd}
    \{T^k_\alpha,m^k_\alpha,V^k_{\alpha;1,\cdots,k}\} = \mathrm{argmin}_{TmV}\delta^k_\alpha,
\end{equation}
with additional smoothness constraints. As $k$ increases, $\{T^k_\alpha,m^k_\alpha,\Tilde{X}^k_{\alpha\nu}\}$ are expected to approach $\{T_\alpha,m_\alpha,X_{\alpha\nu}\}$. In practice, small $k$ is used for ease of computation, and hence $\delta^k_\alpha$ also contains considerable textures as $\delta^k_\alpha \propto \sum_{\beta\neq 1,2,\cdots,k}V_{\alpha\beta}S_{\beta\nu}$. Notably, if those $k$ equivalent environmental objects do not include sky, the whole term of $\Tilde{X}^k_{\alpha\nu}$ needs to be removed from $X_{\alpha\nu}$ to mimic daylight optical imaging, as explained in \sref{sec:gttex}. $\delta^k_\alpha \propto ||\Tilde{X}^k_{\alpha\nu} - X_{\alpha\nu} || $ is exactly the right quantity to find in order to show textures. When sky is included in the equivalent environmental objects, we first inversely solve $\{T^k_\alpha,m^k_\alpha,V^k_{\alpha;1,\cdots,k}\}$ from the heat signal according to \eref{eq:sgd}. Then we forwardly evaluate the iterative equations (\ref{eq:heatsignal}) and (\ref{eq:x}) with $\{T^k_\alpha,m^k_\alpha\}$ but keeping only the sky thermal lighting factor $V^k_{\alpha;\mathrm{sky}}$ to get a distilled texture $\Bar{X}_\alpha$. Finally, the distilled texture $\Bar{X}_\alpha$ is fused with $\delta^k_\alpha$ to get the final texture used in TeX vision.

\begin{figure}[htb]
\centering
\includegraphics{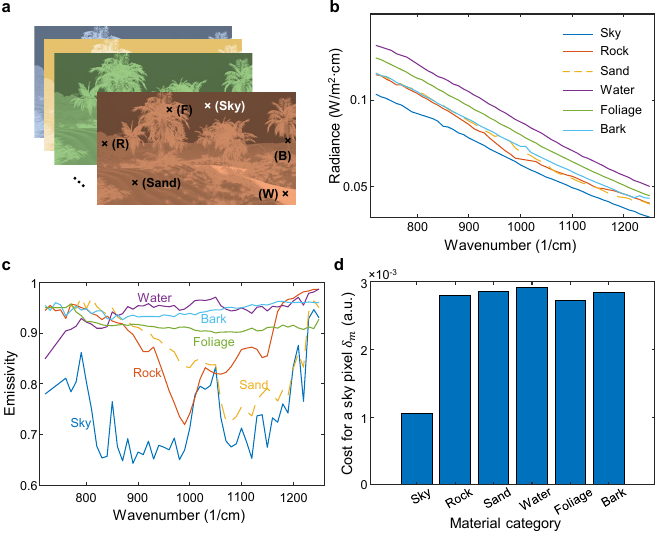}
\caption{Workflow of TeX-SGD. (a) The hyperspectral heat cube. (b) Sample spectral radiance curves for the pixels cross-marked in (a). (c) The material library $\calM$ used for TeX-SGD. This material library is the ground truth library that we used to simulate the scene. (d) The residues in fitting the radiance curve of the pixel marked `Sky' with all possible materials in the library. The minimum residue gives a prediction of `Sky', which is correct. In (a), B: Bark, F: Foliage, R: Rock, and W: Water.}
\label{fig:sgdflow}
\end{figure}

\fref{fig:sgdflow} illustrates the process to solve $\{T^k_\alpha,m^k_\alpha,V^k_{\alpha;1,\cdots,k}\}$ for a given sample sky pixel, with $k=2$.
The sky pixel is marked in \fref{fig:sgdflow}a as a white cross. \fref{fig:sgdflow}b shows sample spectral radiance curves for different materials. The subtle difference in the spectral radiance signal is crucial to distinguishing the materials. \fref{fig:sgdflow}c shows the ground truth material library we used for the TeX decomposition. In practice, we define $\delta_m = \mathrm{min}_{TV}\delta^k_\alpha$ for each pixel $\alpha$. \eref{eq:sgd} therefore gives $m^k_\alpha = \mathrm{argmin}_{m}\delta_m$. \fref{fig:sgdflow}d shows $\delta_m$ for the sample sky pixel. The minimum residue correctly predicts it as `Sky'. Continuous parameters of temperature $T^k_\alpha$ and thermal lighting factors $V^k_{\alpha;1,\cdots,k}$ can be readily solved out by minimizing $\delta_\mathrm{sky}$ using nonlinear least-squares algorithms. The above procedures are repeated for each pixel for a local decomposition. We then impose a smoothness penalty on the resulting $T^k_\alpha$ and $m^k_\alpha$, in addition to $\delta^k_\alpha$, for a global decomposition.

\begin{figure}[htb]
\centering
\includegraphics{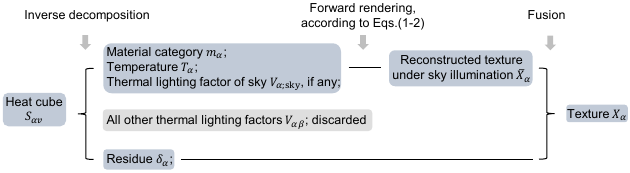}
\caption{Workflow of texture distillation in TeX vision, to reconstruct the texture keeping only the sky illumination, mimicking daylight optical imaging.}
\label{fig:distill}
\end{figure}

\fref{fig:distill} shows the TeX vision process to reconstruct texture under sky illumination, mimicking daylight optical imaging. Note that for the estimated TeX vision, the inverse decomposition is based on TeX-SGD. For the ground truth TeX vision, the Monte Carlo path tracing simulation generates the ground truth decomposition.

\begin{figure}[htb]
\centering
\includegraphics{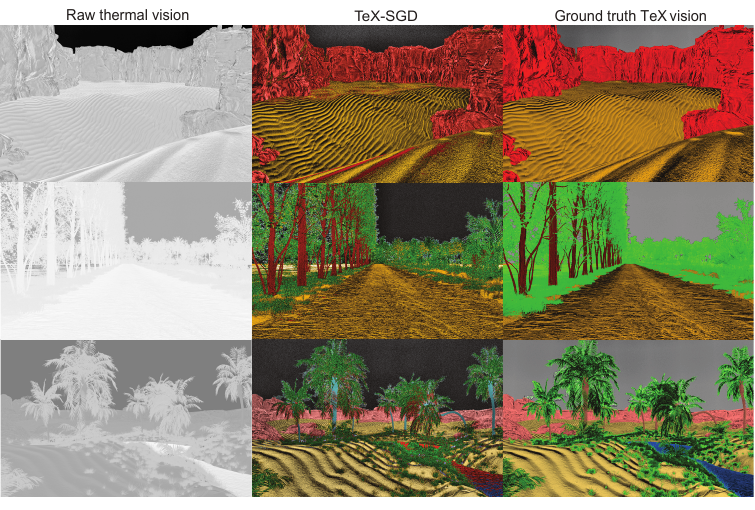}
\caption{TeX vision generated by TeX-SGD recovers geometric textures and disentangles temperature and emissivity close to the ground truth.}
\label{fig:sgd}
\end{figure}

\fref{fig:sgd} shows the TeX vision generated by TeX-SGD, in comparison with the raw thermal vision and the ground truth TeX vision explained in \sref{sec:gttex}.
We emphasize that TeX-SGD has not been previously tested on the synthetic scenes in the HADAR database with ground truth. \fref{fig:sgd} clearly shows the TeX vision generated by TeX-SGD according to \eref{eq:sgd} has recovered the geometric textures as well as semantic information close to the ground truth. However, errors also exist due to the inaccurate approximation of the environmental radiance. See \sref{sec:env} for more analyses.

\section{Interplay of geometric textures and temperature variation}\label{sec:tmp}
For scenes with artificially designed uniform temperatures, texture recovery may be possible by traditional image processing. Furthermore, it is unclear if TeX vision theory can recover textures for common scenes with non-uniform temperatures. Here, we study realistic scenes with non-uniform temperatures, in addition to geometric surface normals, and show the advantage of TeX vision with respect to traditional image processing. Differentiating \eref{eq:heatsignal}, we have
\begin{equation}\label{eq:alltex}
    \delta S = \delta T \cdot e\partial_{T}B + \delta e\cdot [B(T) - X] + \delta X\cdot(1-e).
\end{equation}
Here, we have suppressed the subscripts for clarity. The overall signal variation in the image consists of 3 contributions: the material change $\delta e$, the temperature contrast $\delta T$, and geometric texture $\delta V$ in the term $\delta X = \delta \vec{V}\cdot \vec{S} + \delta \vec{S}\cdot \vec{V}$. \eref{eq:alltex} indicates that recovering the geometric textures requires the separation of the 3 variation contributions. We argue this is possible with spectral resolution, but in principle, impossible for traditional thermal imaging. This inseparability of geometric textures with temperature and material variations in panchromatic thermal imaging is the fundamental reason underlying the ghosting effect.

\begin{figure}[htb]
\centering
\includegraphics{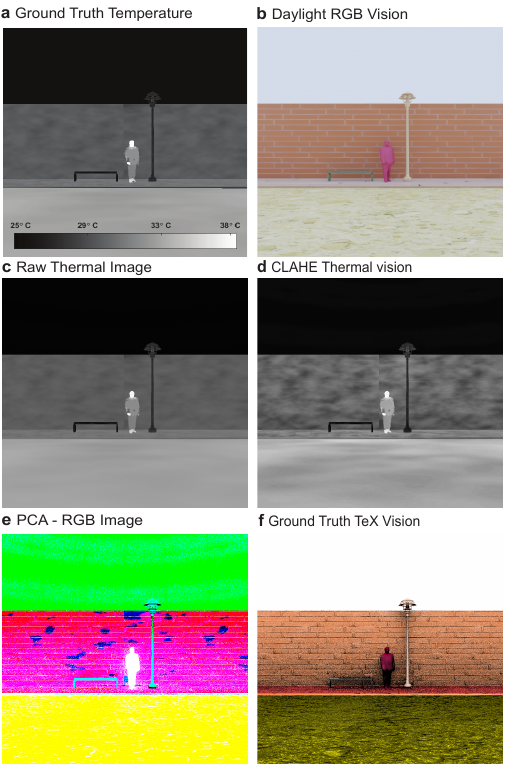}
\caption{TeX vision recovers geometric textures from non-uniform temperature contrast, while traditional thermal imaging with image processing fails. The interplay of temperature contrast and geometric textures makes it difficult for thermal imaging to recover the geometric texture, and this interplay is the fundamental reason causing the ghosting effect. (a) Ground truth temperature. (b) Optical imaging in daylight with geometric textures. (c) Raw thermal imaging at night. (d) State-of-the-art CLAHE thermal vision. (e) State-of-the-art principle component analysis. (f) Ground truth TeX vision.}
\label{fig:tmp}
\end{figure}

\fref{fig:tmp} shows a newly designed scene with both temperature contrast and geometric textures within each object.
The ground truth temperature in \fref{fig:tmp}a and the optical imaging in \fref{fig:tmp}b show the temperature contrast and vivid geometric textures, respectively. However, geometric textures become invisible after being immersed in the temperature contrast using only raw thermal imaging, see \fref{fig:tmp}c. The state-of-the-art CLAHE algorithm can improve the visual contrast but fails to separate the geometric textures from temperature contrast, see \fref{fig:tmp}d. Furthermore, the state-of-the-art principle component analysis approach also fails to separate the geometric textures from temperature contrast, see \fref{fig:tmp}e. On the contrary, TeX vision recovers the geometric textures, enabling a night vision that approaches the texture contrast present in daylight optical imaging. This result clearly demonstrates the advantage of TeX vision in overcoming the ghosting effect for realistic scenes with non-uniform temperatures. See \fref{fig:tmp}f for the ground truth TeX vision and \fref{fig:4band} for the TeX vision generated by TeX-SGD.

\section{Influence of spectral resolution}\label{sec:res}
\begin{figure}[htb]
\centering
\includegraphics{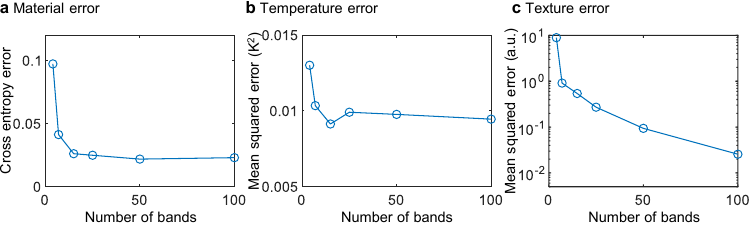}
\caption{Prediction error of TeX-SGD decreases with increasing spectral resolution. (a) Cross entropy error for material classification. (b) Mean squared error of temperature estimation. (c) Mean squared error of texture estimation normalized by the number of bands. 5 equivalent environmental objects were used by K-means clustering.}
\label{fig:res}
\end{figure}

Spectral resolution plays a vital role in solving the inverse problem of \eref{eq:sgd}. In the absence of spectral resolution, thermal camouflage \cite{Bao2023,Li2020,Qu2018June} leads to ambiguous solutions. How many spectral bands are needed for a TeX vision generally depends on the specific scene and how many materials we want to discern. Here, we study the error scaling law of TeX-SGD with respect to the ground truth TeX vision for various spectral resolutions. The analysis is based on the scene shown in \fref{fig:tmp}. The scene was rendered with 100 spectral bands equidistantly distributed within the LWIR (8-14\,$\mu$m). Out of the 100 bands, we choose an equidistant subset to generate TeX vision and derive the error of predicted material and temperature. \fref{fig:res} shows the error scaling law as a function of the number of spectral bands. Our results confirm that prediction errors of TeX-SGD decrease with increasing spectral resolution.

\begin{figure}[htb]
\centering
\includegraphics{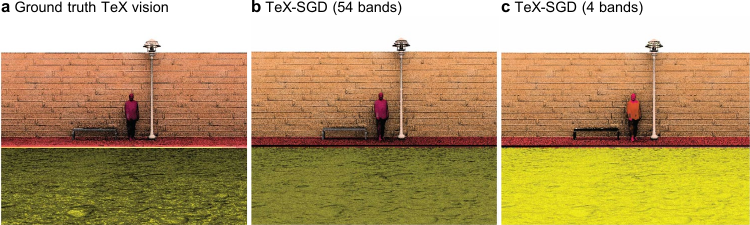}
\caption{TeX vision in the Bayer-filter approach shows the experimental feasibility of applying TeX vision in real-time applications.}
\label{fig:4band}
\end{figure}

LWIR hyperspectral imaging is experimentally very challenging and expensive. Multi-spectral imaging by the Bayer-filter approach \cite{BRINEZDELEON2019195} using 4 filters can operate in snapshot mode, and hence is promising for real-time applications such as autonomous navigation. \fref{fig:4band} demonstrates the TeX vision of TeX-SGD based on 4 spectral bands in comparison with the ground truth TeX vision and the TeX vision based on 54 bands. The fact that 4-band TeX vision can recover geometric textures and even come close to the ground truth TeX vision shows the possibility of implementing TeX vision using the Bayer-filter approach, which is much more feasible in experiments than LWIR hyperspectral imaging.

\section{Cutoff on the number of environmental objects}\label{sec:env}
Here we analyze the dependence of TeX vision on the approximation of environmental radiance. According to Eqs.~(\ref{eq:approxheat}) and (\ref{eq:approxx}), the accuracy of the predicted TeX vision depends on the modeling of the environmental radiation. A panoramic image is ideally needed to correctly characterize the environment. If the field of view is restricted, the captured image (\ie, the spectral data cube) is used to approximate significant environmental objects.

\begin{figure}[htb]
\centering
\includegraphics{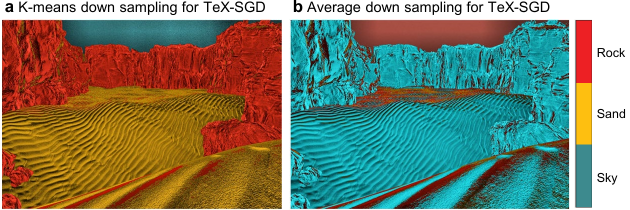}
\caption{The influence of down sampling methods on the accuracy of TeX vision.}
\label{fig:dn}
\end{figure}

\fref{fig:dn} shows the TeX vision comparison, with average down sampling vs. the K-means down sampling, for two equivalent environmental objects ($k=2$). For average down sampling, we split the spectral data cube (Image height $\times$ Image width $\times$ Spectral bands) into upper and lower halves along the height direction, spatially average each sub data cube, and get two radiation spectra. The upper spectrum approximates the sky radiation, while the lower spectrum approximates the ground radiation. In K-means down sampling, we perform K-means clustering with $k=2$ on all image pixels, and each cluster is spatially averaged to get the spectrum. As can be seen, K-means down sampling for environment approximation shows more accurate TeX vision. This is reasonable, as K-means clustering is better than coarse meshing at extracting objects with irregular shapes.

\begin{figure}[htb]
\centering
\includegraphics{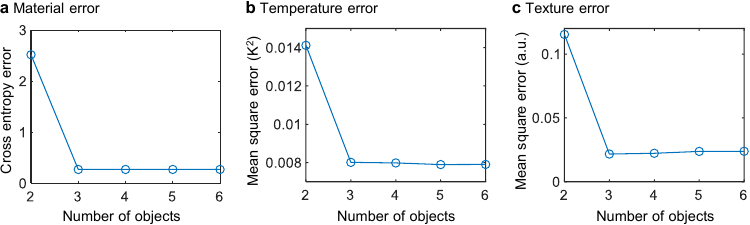}
\caption{Error in predicted TeX vision as a function of the number of equivalent environmental objects. (a) Cross entropy error for material classification. (b) Mean squared error of temperature estimation. (c) Mean squared error of texture estimation normalized by the number of bands. 54 spectral bands were used.}
\label{fig:env}
\end{figure}

We used K-means down sampling for all experiments unless otherwise specified. \fref{fig:env} further shows the error in predicted TeX vision for various preset numbers of equivalent environmental objects ($k$). $k$ is used as the input to K-means clustering.
Due to specific orientations, every object has its own set of environmental objects that dominate its scattering signal. It follows that more environmental objects (\ie, larger $k$) give lower error in solving \eref{eq:sgd}. However, larger $k$ immediately results in more variables of thermal lighting factors $V_k$, which in turn requires higher spectral resolution to solve the problem. This explains, for a given spectral resolution, why errors steadily approach constants in \fref{fig:env}.

\section{Conclusion}
We studied the interplay of geometric textures and non-uniform temperature contrast in thermal imaging. We have shown that traditional thermal imaging lacking spectral resolution cannot resolve geometric textures from temperature variation, and we argue this is the fundamental reason causing the ghosting effect. This work verifies the TeX vision theory and demonstrates its advantage over traditional thermal imaging in overcoming the ghosting effect. Furthermore, we have demonstrated TeX vision with the Bayer-filter approach with low spectral resolution. This relieves the experimental challenge of TeX vision, enabling its real-time applications in, for example, autonomous navigation, robotics, wildlife monitoring, smart healthcare, geoscience, and defense.

\begin{backmatter}
\bmsection{Funding}
This work was supported by the Invisible Headlights project from the Defense Advanced Research Projects Agency (DARPA).

\bmsection{Disclosures}
The authors declare no conflicts of interest.

\bmsection{Data Availability Statement}
Data underlying the results presented in this paper are available at \url{https://github.com/FanglinBao/HADAR}.

\end{backmatter}

\bibliography{tex}




\end{document}